\documentclass[iop]{emulateapj}
\usepackage{amsmath}
\usepackage{amssymb}
\usepackage{verbatim}
\usepackage{graphicx}
\usepackage{morefloats}
\usepackage{float}
\usepackage{lipsum}
\usepackage{lipsum} 
\usepackage{rotating}
\usepackage{url} 

\begin{document}

\title{SuperNova IDentification spectral templates of 70 stripped-envelope core-collapse supernovae}

\author{Yuqian Liu and Maryam Modjaz}
\thanks{Yuqian Liu: yl1260@nyu.edu \\Maryam Modjaz: mmodjaz@nyu.edu}
\affil{Center for Cosmology and Particle Physics, New York University, New York, NY 10003}

\begin{abstract}
We constructed 70 SuperNova IDentification \citep*[SNID;][]{2007ApJ...666.1024B} supernova (SN) templates using 640 spectra of stripped-envelope core-collapse SNe (SESNe) published by \citet*{2014AJ....147....99}. Fifty-six SN templates which are constructed from 458 spectra are brand new, increasing the number of SESNe and the number of SESNe spectra in the current SNID database by a factor of 2.6 and 2.0, respectively. We also made some type and phase refinements to templates in the current SNID database.
\end{abstract}

\section{Introduction}
The SuperNova IDentification code (SNID\footnote[1]{http://people.lam.fr/blondin.stephane/software/snid/index.html}; \citealt{2007ApJ...666.1024B}) classifies a supernova (SN) spectrum by cross-correlating the spectrum with template spectra in its database. These template spectra are continuum-removed and de-redshifted spectra with known type and phase with respect to either the date of first observation or the date of maximum light. Template spectra of the same SN form a SN template. The latest version of the SNID database (templates-2.0$^1$) contains 465 spectra of 36 Type Ib/c SNe (SNe Ib/c)\footnote[2]{While the class of stripped-envelope core-collapse SNe (SESNe; \citealt{1997ApJ...483..675C}) includes SNe of types IIb, Ib, Ic and Ic-bl, to be consistent with the SNID nomenclature, we will subdivide them broadly into Type Ib (which includes Ib-norm, IIb, Ib-pec and Ib-n as subtypes) and Type Ic (which includes Ic-norm and Ic-broad as subtypes).}, 2724 spectra of 283 Type Ia SNe (SNe Ia), 527 spectra of 14 Type II SNe (SNe II) and some spectra of non-SN objects (active galactic nuclei, galaxies, luminous blue variables and M-stars). Fourteen out of the 36 SN Ib/c templates in templates-2.0 were constructed mainly using spectra that were published by \citet*[hereafter M14]{2014AJ....147....99}. While SNID is a widely used classification code, its current database (templates-2.0) has room for improvement in at least two aspects. One aspect is to increase the number of SN Ib/c templates. The number of SNe Ib/c in templates-2.0 is much smaller than that of SNe Ia. As shown in \citet*{2007ApJ...666.1024B}, this may lead to a type ``attractor'' that increases the risk of misclassifying a SN Ib/c spectrum with a low signal to noise ratio (S/N) as a SN Ia spectrum because there are more SN Ia spectra for the input spectrum to cross-correlate with. The second aspect is to update the type and phase information of some SNe Ib/c in templates-2.0 since M14 found that up to 1/3 of their SN sample had a different type than announced in the SN catalogs or in the International Astronomical Union Circulars (IAUC). \citet*{2012MNRAS.425.1789S} created Berkeley Supernova Ia Program (BSNIP) v7.0 templates which mainly focus on SNe Ia and have no phase information for SNe Ib/c.  

M14 recently published 645 optical spectra of 73 stripped-envelope core-collapse SNe (SESNe; \citealt{1997ApJ...483..675C}) collected at the Harvard-Smithsonian Center for Astrophysics (CfA). The data are homogenous given that they were mostly collected with the same instrument on the same telescope and were reduced in the same manner. Using these spectra, we created CfA spectral templates for SNID use which can be downloaded via our SNYU webpage\footnote[3]{http://cosmo.nyu.edu/SNYU/spectra}. They will improve the current SNID database in the above two aspects.

\section{CfA spectral templates}

\subsection{A new SNID subtype}

The current version of SNID (SNID-5.0$^1$) divides SNe Ib/c into the following subtypes: Ib-norm, Ib-pec, IIb, Ic-norm, Ic-pec and Ic-broad. We introduced the subtype Ib-n \citep*{2007Natur.447..829P} to SNID by modifying the code \textit{typeinfo.f}\footnote[4]{http://people.lam.fr/blondin.stephane/software/snid/faq.html-\#newtype} in the SNID source subdirectory. SNe of this new subtype are dominated by narrow He I lines in their spectra, which are due to interaction of the SN ejecta with a He-rich circumstellar medium around their progenitor stars \citep{2000AJ....119.2303M, 2007ApJ...657L.105F, 2008MNRAS.389..113P}. We introduced this new subtype since we included the CfA spectra of SN 2006jc---the most well studied SN Ib-n and the only SN Ib-n for which we have CfA data although there are seven more SNe Ib-n identified so far in the literature---into the spectral templates. Although there is the subtype ``Ic-pec'' in the current SNID classification scheme, there is no SN Ic-pec template in either the templates-2.0 release or in our CfA templates. This subtype was used to represent SNe Ic-broad when people knew little about this new kind of SNe.

\subsection{New SNID templates}

We created SNID templates using 640 CfA spectra of the 70 SESNe published by M14. Generated by the program \textit{logwave}\footnote[5]{http://people.lam.fr/blondin.stephane/software/snid/howto.html-\#logwave}, they have the same format as templates in templates-2.0 and can be used in SNID-5.0. SN 1995bb, SN 2005la and SN 2007iq from M14 are not included in the CfA spectral templates because their types are uncertain. Some spectra from the paper are not used due to their low-S/N. Telluric lines were removed during data reduction, following a procedure similar to that of \citet*{1988ApJ...324..411W} and \citet*{2000AJ....120.1487M} . We used simple interpolation to remove emission lines from H II regions whenever they contaminated the spectra. Table \ref{table_summary} shows a summary of the CfA SN type distribution. For 43 SNe of the 70 SNe in the CfA SN templates, we have dates of maximum light either from Bianco et al. (2014, ApJS submitted) or from other literature sources (see table 2 in M14). Table \ref{table_nomax} and table \ref{table_max} show the subtype and rounded phase of each CfA spectral template. Fifty-six SNe among the CfA SN templates are new, enlarging the number of SNe Ib/c in templates-2.0 by 160\% and the number of SNe Ib/c spectra by 100\%. The remaining 14 SNe (see table 4) in the SNID templates-2.0 release were constructed by \citet*{2007ApJ...666.1024B} who used both CfA spectra before being published as such in M14 and spectra from other literature sources (see their table 1). In Table 4 we comment on each of these SNe and give recommendations on whether SNID users should adopt our CfA SN template or stick to the original SNID template in the templates-2.0 release. Six of them have updated type or phase information according to M14, as is shown in table \ref{table_update}. Finally, we have updated the type of SN 1990U, which is a part of the templates-2.0 release but is not a part of the CfA SN templates, from Ic-norm to Ib-norm (see table 3 in M14).

\acknowledgments
We are grateful to Stephane Blondin, Federica Bianco and Or Graur for useful discussions and for comments on this manuscript.

Y. Liu is in part supported by a NYU/CCPP James Arthur Graduate Award. M. Modjaz is supported in parts by the NSF CAREER award AST-1352405.

\begin{deluxetable}{cc}
\tablecolumns{2}
\singlespace
\tablecaption{CfA SN templates type distribution}
\tablehead{
\colhead{(Sub)Type} &
\colhead{No. of SNe}  
}
\startdata
Ib (total)  & 40   \\     
Ib-norm  &  24     \\
Ib-pec  &  2       \\
IIb  &  13         \\
Ib-n  &  1          \\
\hline
Ic (total)  & 30    \\ 
Ic-norm    &  19    \\
Ic-broad    & 11    \\
\enddata
\label{table_summary}
\end{deluxetable}

\newpage 
\bibliographystyle{plain}

\begin{deluxetable*}{llc}
\tablecolumns{3}
\singlespace
\tablecaption{Summary of the CfA spectral templates without a date of maximum light}
\tablehead{
\colhead{SN Name} &
\colhead{Subtype} &
\colhead{Phases (wrt date of first observation)\tablenotemark{a}}  
}
\startdata
SN 1995F & Ib-norm &  0, 2, 4, 10,
30+(1) \\ [1ex]
SN 1997X & Ib-norm& 0, 2, 5, 7, 10,
25\\ [1ex]
SN 1997dq & Ic-norm &  0, 15, 17, 21, 26, 27,
30+(1) \\ [1ex]
SN 2001ai & Ib-norm& 0,
27\\ [1ex]
SN 2001ej & Ib-norm &  0, 1, 2, 28,
30+(4) \\ [1ex]
SN 2001gd & IIb &  0,
30+(1) \\ [1ex]
SN 2002ji & Ib-norm &  0, 5, 8, 21, 23, 27, 30,
30+(5) \\ [1ex]
SN 2004ao & Ib-norm &  0, 1, 2, 3, 4, 5, 6, 8, 9, 12, 13, 14, 15, 30,
30+(16) \\ [1ex]
SN 2004eu & Ic-norm\tablenotemark{b}& 0, 2, 4, 5, 9,
10\\ [1ex]
SN 2004gk & Ic-norm &  0, 6, 7, 9, 10, 11, 12, 13, 15, 17, 18,
30+(19) \\ [1ex]
SN 2005U & IIb &  0, 28,
30+(2) \\ [1ex]
SN 2005ar & Ib-norm& 0,
2\\ [1ex]
SN 2005da & Ic-bl& 0, 2,
3\\ [1ex]
SN 2005kf & Ic-norm& 0, 3, 7, 7,
12\\ [1ex]
SN 2005nb & Ic-bl& 0
\\ [1ex]
SN 2006ck & Ic-norm& 0, 1, 2, 2, 4, 5, 6,
7\\ [1ex]
SN 2006fo & Ib-norm &  0, 1,
30+(1) \\ [1ex]
SN 2006lc & Ib-norm& 0, 1, 3,
4\\ [1ex]
SN 2006ld & Ib-norm& 0
\\ [1ex]
SN 2006lv & Ib-norm &  0, 3, 5, 8, 11,
30+(1) \\ [1ex]
SN 2007I & Ic-bl &  0, 4, 9,
30+(1) \\ [1ex]
SN 2007ce & Ic-bl& 0, 1, 3, 5, 7, 9, 11, 15,
30\\ [1ex]
SN 2007hb & Ic-norm &  0,
30+(1) \\ [1ex]
SN 2007rz & Ic-norm &  0,
30+(1) \\ [1ex]
SN 2008an & Ic-norm& 0,
1\\ [1ex]
SN 2008aq & IIb &  0, 1, 2, 3, 9,
30+(3) \\ [1ex]
SN 2008cw & IIb& 0, 1, 1, 3,
6\\ [1ex]
\enddata
\tablenotetext{a}{Phases are in rest-frame and rounded to the nearest whole day. Number in a bracket is the number of spectra with phases larger than 30 days after the date of first observation.}
\tablenotetext{b}{M14 couldn't rule out the potential possibility of the emergence of He I lines in these SNe Ic, since their spectra were taken either before or well after maximum light, when the He lines are not as well pronounced.}

\label{table_nomax}
\end{deluxetable*}

\begin{deluxetable*}{llc}
\tablecolumns{3}
\singlespace
\tablecaption{Summary of the CfA spectral templates with a date of maximum light from table 2 in M14}
\tablehead{
\colhead{SN name} &
\colhead{Subtype} &
\colhead{Phases (wrt date of maximum light)\tablenotemark{a}}  
}
\startdata
SN 1993J & IIb & 30+(10) \\ [1ex]
SN 1994I & Ic-norm &  -5, -5, -3, -2, -1, 0, 1, 2, 2, 3, 21, 22, 23, 24, 26, 28, 30,
30+(6) \\ [1ex]
SN 1996cb & IIb &  -19, -18, -3, 0, 3, 4, 5, 6, 25, 27,
30+(12) \\ [1ex]
SN 1997ef & Ic-bl &  -15, -13, -12, -11, -10, -7, -6, -6, -5, 11, 12, 14, 15, 17, 18, 20, 22, 25,
30+(7) \\ [1ex]
SN 1998dt & Ib-norm &  0, 1, 3, 6, 10, 11,
16 \\ [1ex]
SN 1998fa & IIb &  -5, -4,
16 \\ [1ex]
SN 2000H & IIb &  -2, -1, 1, 2, 3, 14,
16 \\ [1ex]
SN 2002ap & Ic-bl &  -1, 0, 0, 1, 4, 5, 6, 7, 12, 26, 30,
30+(4) \\ [1ex]
SN 2003jd & Ic-bl &  -2, -1, 0, 0, 1, 18, 19, 20, 21, 22, 23, 26, 27,
30+(12) \\ [1ex]
SN 2004aw & Ic-norm &  -5, -3, -3, -2, -1, 0, 1, 2, 3, 17, 20, 25, 29,
30+(2) \\ [1ex]
SN 2004dk & Ib-norm &  17, 20,
30+(1) \\ [1ex]
SN 2004dn & Ic-norm &  30,
30+(3) \\ [1ex]
SN 2004fe & Ic-norm &  -6, -6, -5, -4, 0, 0, 1, 2, 3, 9, 21,
30+(1) \\ [1ex]
SN 2004ff & IIb &  -3
 \\ [1ex]
SN 2004ge & Ic-norm &  5
 \\ [1ex]
SN 2004gq & Ib-norm &  -9, -8, -7, -6, -5, -2, -1, 0, 15, 16, 17, 20, 20, 24,
30+(10) \\ [1ex]
SN 2004gt & Ic-norm &  16, 18, 19, 22,
30+(9) \\ [1ex]
SN 2004gv & Ib-norm &  13, 15, 18,
30+(1) \\ [1ex]
SN 2005az & Ic-norm &  -8, -7, -5, -3, -1, 0, 3, 16, 18, 21, 23, 25, 27, 29, 30,
30+(7) \\ [1ex]
SN 2005bf & Ib-norm &  -29, -27, -26, -25, -23, -8, -6, -5, -4, -3, -2, -1, 0, 1, 2, 4, 5, 7, 22, 25, 29,
30+(2) \\ [1ex]
SN 2005hg & Ib-norm &  -13, -12, -10, -9, -8, -7, -6, -5, -4, -3, -2, -1, 0, 11, 15, 21, 25,
30+(1) \\ [1ex]
SN 2005kl & Ic-norm\tablenotemark{b} &  -4,
30+(2) \\ [1ex]
SN 2005mf & Ic-norm &  -2, -1, 4,
7 \\ [1ex]
SN 2006T & IIb &  -13, -11, 6, 14, 21,
30+(2) \\ [1ex]
SN 2006aj & Ic-bl &  -6, -5, -4, -3, -2, -1, 0, 1, 2,
30+(1) \\ [1ex]
SN 2006el & IIb &  -4, 9, 10, 11, 12, 15,
16 \\ [1ex]
SN 2006ep & Ib-norm &  -8, 7, 9, 11,
30+(1) \\ [1ex]
SN 2006jc & Ib-n &  5, 7, 8, 9, 10, 11, 15, 17, 19, 30,
30+(9) \\ [1ex]
SN 2007C & Ib-norm &  -6, -5, -1, 0, 6, 27,
30+(6) \\ [1ex]
SN 2007D & Ic-bl &  -7
 \\ [1ex]
SN 2007ag & Ib-norm &  0,
9 \\ [1ex]
SN 2007bg & Ic-bl &  3
 \\ [1ex]
SN 2007cl & Ic-norm\tablenotemark{b} &  -6
 \\ [1ex]
SN 2007gr & Ic-norm &  7, 8, 9, 10, 13, 14, 15, 16, 17, 18, 20, 25,
30+(6) \\ [1ex]
SN 2007kj & Ib-norm &  -1,
3 \\ [1ex]
SN 2007ru & Ic-bl &  -3, 0,
30+(1) \\ [1ex]
SN 2007uy & Ib-pec &  -13, -11, -6, -5, -3, 12,
30+(4) \\ [1ex]
SN 2008D & Ib-norm &  -17, -16, -15, -14, -13, -12, 3, 10, 13,
30+(1) \\ [1ex]
SN 2008ax & IIb &  -20, -19, -18, -16, -15, -13, -12, -11, -10, -9, -8, 6, 8, 9, 9, 12, 15, 16, 19, 23,
30+(6) \\ [1ex]
SN 2008bo & IIb &  -10, -7, -3, -1, 15, 20, 24, 30,
30+(4) \\ [1ex]
SN 2009er & Ib-pec &  -4, -4, -2, -1, 0, 2, 15,
16 \\ [1ex]
SN 2009iz & Ib-norm &  -13, -9, 5, 10, 11, 12, 20,
30+(2) \\ [1ex]
SN 2009jf & Ib-norm &  -8, -4, -1, 0, 1, 7, 23, 28,
30+(8) \\ [1ex]
\enddata
\tablenotetext{a}{Phases are in rest-frame and rounded to the nearest whole day. Number in a bracket is the number of spectra with phases larger than 30 days after the date of maximum light. All dates of maximum light were measured in the $V$-band, except for those of SN 1998dt, SN 1998fa, SN 2004ff, SN 2004ge, and SN 2007bg, which were measured in the $R$-band.}
\tablenotetext{b}{M14 couldn't rule out the potential possibility of the emergence of He I lines in these SNe Ic, since their spectra were taken either before or well after maximum light, when the He lines are not as well pronounced.}
\label{table_max}
\end{deluxetable*}

\begin{deluxetable*}{lcc}
\tablecolumns{3}
\singlespace
\tablecaption{Summary of 14 SN templates that are in both the templates-2.0 and the CfA SN templates}
\tablehead{
\colhead{SN name} &
\colhead{No. of spectra in the templates-2.0 release} &
\colhead{No. of spectra in the CfA spectral templates} 
}
\startdata
SN 1993J \tablenotemark{a} &  74     &      10 \\ [0.5ex]
SN 1994I \tablenotemark{b} &    
         24     &      23\\ [0.5ex]
SN 1995F \tablenotemark{b}\tablenotemark{c} &
          10     &      5 \\ [0.5ex]
SN 1996cb \tablenotemark{b}\tablenotemark{c} &
          25      &     22 \\ [0.5ex]
SN 1997dq \tablenotemark{b}\tablenotemark{c} &
          12     &      7\\ [0.5ex]
SN 1997ef \tablenotemark{b} &
          28     &      25\\ [0.5ex]
SN 1998dt \tablenotemark{b} & 
           9      &      7\\ [0.5ex]
SN 2000H \tablenotemark{b}\tablenotemark{c} &
          12      &      7\\ [0.5ex]
SN 2002ap \tablenotemark{b} &
          34     &      15 \\ [0.5ex]
SN 2004aw \tablenotemark{b} &
          27      &     15 \\ [0.5ex]
SN 2005bf \tablenotemark{b}\tablenotemark{c} &
          30       &    23 \\ [0.5ex]
SN 2005kl \tablenotemark{d} &
           1       &     3 \\ [0.5ex]
SN 2006aj \tablenotemark{d} &
           8      &     10 \\ [0.5ex]
SN 2008D \tablenotemark{b}\tablenotemark{c} &
          21      &     10 \\ [0.5ex]
\enddata
\tablenotetext{a}{For this SN, its CfA spectra were taken at more than 590 days after the date of maximum light. We suggest that SNID users do not overwrite its template in the templates-2.0 release with the one in the CfA SN templates.}
\tablenotetext{b}{For this SN, we suggest that SNID users do not overwrite its template in the templates-2.0 release with the one in the CfA SN templates because the original SNID template contains both our CfA spectra and spectra from other sources \citep*[see table 1 in][]{2007ApJ...666.1024B}.}
\tablenotetext{c}{For this SN, if SNID users want to use its original SNID template, they should update its type or phase information according to table \ref{table_update} below.}
\tablenotetext{d}{For this SN, we suggest that SNID users overwrite its template in the templates-2.0 release with the one in our CfA SN templates because the latter includes not only all spectra that were used to construct the original SNID template but also additional CfA spectra.}
\label{table_common}
\end{deluxetable*}

\begin{deluxetable*}{ll}
\tablecolumns{2}
\singlespace
\tablecaption{Summary of 6 SN templates that are updated in the CfA SN templates compared with those in the templates-2.0 release}
\tablehead{
\colhead{SN name} &
\colhead{comments} 
}
\startdata
SN 1995F  &   the type is changed from Ic-norm to Ib-norm        \\ [0.5ex]
SN 1997dq &   the type is changed from Ic-broad to Ic-norm   \\ [0.5ex]
SN 1996cb &   the MJD date of maximum light is changed from 50452.5 to 50453.9\\ [0.5ex]
SN 2000H & the MJD date of maximum light is changed from 51577.0 to 51585.0\tablenotemark{a} \\ [0.5ex]
SN 2005bf  &  the type is changed from Ib-pec to Ib-norm \\ [0.5ex]
SN 2008D  &  the type is changed from Ib-pec to Ib-norm; the MJD date of maximum light is changed from 54492.9 to 54494.1 \\ [0.5ex]
\enddata
\tablecomments{For a complete list of SN type refinements, see table 3 in M14. SNe with updated dates of maximum light in this table have a $>1$-day difference between the original value reported in the templates-2.0 release and the one computed in M14.}
\tablenotetext{a}{SN 2000H has an uncertain date of maximum light: while Branch et al. (2002) favor 2000 February 11 as the date of V-band max from unpublished ESO photometry, the Asiago Catalogue lists 2000 February 2 as the date of max. Here we adopt the later date, consisting with Branch et al. (2002).}
\label{table_update}
\end{deluxetable*}

\end{document}